\documentclass[aps,prl,amsmath,amssymb,twocolumn,floatfix,superscriptaddress]{revtex4-1}
\usepackage{graphicx}
\usepackage{dcolumn}
\usepackage{bm}
\usepackage{subfigure}
\usepackage{amsmath}
\usepackage{hyperref}

\usepackage{times}

\newcommand{\ua}{\uparrow}
\newcommand{\da}{\downarrow}
\newcommand{\ra}{\rightarrow}

\newcommand{\bs}{\boldsymbol}

\newcommand{\SRO}{Sr$_2$RuO$_4$}

\begin{document}
\title{Possible three-dimensional nematic odd-parity superconductivity in \SRO}
\author{Wen Huang}
\affiliation{Institute for Advanced Study, Tsinghua University, Beijing, 100084 China}
\author{Hong Yao}
\email{yaohong@tsinghua.edu.cn}
\affiliation{Institute for Advanced Study, Tsinghua University, Beijing, 100084 China}
\affiliation{State Key Laboratory of Low Dimensional Quantum Physics, Tsinghua University, Beijing 100084, China}
\affiliation{Collaborative Innovation Center of Quantum Matter, Beijing 100084, China}
\date{\today}

\begin{abstract}
The superconducting pairing in \SRO~is widely considered to be chiral $p$-wave with $\vec{d}_{\bs k} \sim (k_x + ik_y)\hat{z}$, which belongs to the $E_u$ representation of the crystalline $D_{4h}$ group. However, this superconducting order appears hard to reconcile with a number of key experiments. In this paper, based on symmetry analysis we discuss the possibility of odd-parity pairing with inherent three-dimensional (3D) character enforced by the inter-orbital interlayer coupling and the sizable spin-orbit coupling (SOC) in the material. We focus on a yet unexplored $E_u$ pairing, which contains finite $(k_z \hat x$, $k_z \hat y)$-component in the gap function. Under appropriate circumstances a novel time-reversal invariant nematic pairing can be realized. This nematic superconducting state could make contact with some puzzling observations on \SRO, such as the absence of spontaneous edge current and no evidences of split transitions under uniaxial strains.
\end{abstract}

\maketitle
{\it Introduction}.---Superconductivity in \SRO~was discovered \cite{Maeno:94} in 1994 and was immediately proposed to be of spin-triplet pairing in relation to the possible remnant ferromagnetic correlations in the material \cite{Rice:95,Baskaran:96}. Over the years, multiple measurements show evidences of spin-triplet \cite{Ishida:98,Duffy:00}, odd-parity pairing \cite{Nelson:04}, with the additional feature of time-reversal symmetry breaking (TRSB) \cite{Luke:98,Xia:06}. These point to a possible chiral $p$-wave pairing \cite{Maeno:01,Mackenzie:03,Kallin:09,Kallin:12,Maeno:12,Liu:15,Kallin:16,Mackenzie:17} in the $E_u$ representation, represented by the pairing function $\vec{d}_{\bs k}= (k_x \pm i k_y)\hat z$. Here ``$\pm$" indicate the time-reversed pair of degenerate chiral states, and the direction of the $\vec{d}$-vector, $\hat z$ in this case, denotes the structure of Cooper pairing in spin space (see later). If confirmed, \SRO~will be a solid state analog of the well-known liquid $^3$He A-phase \cite{Vollhardt:90}. This state is topologically nontrivial, wherein the Cooper pairs carry nonvanishing quantized orbital angular momentum. It supports exotic excitations such as chiral edge states and Majorana zero modes in superconducting vortex cores. The latter is marked by non-abelian braiding statistics crucial for topological quantum computation \cite{Kitaev:03,Nayak:08}.

However, the chiral $p$-wave pairing still currently stands in conflict with a number of experimental observations. A prominent example is the absence of spontaneous edge current \cite{Kirtley:07,Hicks:10,Curran:14}. Existing measurements place an upper bound for the edge current over three orders of magnitude smaller than predicted for an isotropic single-band chiral $p$-wave model \cite{Matsumoto:99}. Other inconsistencies include but are not limited to: abundant residual density of states going against the fully-gapped nature of a chiral $p$-wave \cite{Nishizaki:00,Hassinger:16}; signatures reminiscent of the Pauli limiting behavior \cite{Deguchi:02,Rastovski:13,Kuhn:17,Yonezawa:14}; the absence of split transitions in the presence of external perturbations expected to lift the degeneracy of the two $E_u$ components of the chiral order parameter, such as an in-plane magnetic field \cite{Yonezawa:14} and in-plane uni-axial strains \cite{Hicks:14,Steppke:16}, etc.

Recent years have seen a broad spectrum of theoretical attempts to resolve various aspects of the puzzle \cite{Ashby:09,Raghu:10,Sauls:11,Taylor:12,Wysokinski:12,Imai1213,Chung:12,Takamatsu:13,Hughes:14,Wang:13,Huo:13,Bouhon:14,Lederer:14,Huang:14,Huang:15,Tada:15,Tsuchiizu:15,Scaffidi:14,Scaffidi:15,Yada:14,Nakai:15,Amano:15,Sauls:15,Fischer:16,Ramires:16,Huang:16,Cobo:16,Hsu:17,Kim:17,Komendova:17,Zhang:17b,Zhang:17a,Etter:17,Ojanen:16,Suzuki:16, Hsu:16,Liu:17}.  However, a consensus is still lacking regarding the exact pairing symmetry in \SRO. To this end, we take a different angle and study a possible alternative $\vec{d}$-vector in the $E_u$ representation on account of the weak inter-orbital interlayer tunneling and the sizable spin-orbit coupling (SOC) \cite{Haverkort:08,Veenstra:14,Fatuzzo:15} between the Ru 4$d$ $t_{2g}$-orbitals -- which introduce considerable three-dimensional spin-orbit entanglement as reported in photo-emission studies \cite{Veenstra:14}. In addition to the pairing in  the channel ($k_x \hat z, k_y\hat z$), the $\vec{d}$-vector should contain finite $(k_z \hat x, k_z \hat y)$ pairing, thereby constituting a full 3D superconducting pairing. As we shall see, the interplay between these two pairing channels brings about an interesting possibility of a novel time-reversal invariant (TRI) nematic superconducting phase. Such a state is doubly degenerate, possesses symmetry-imposed point-nodal quasiparticle excitations (but could in principle support accidental nodal lines), and exhibits a broken rotational symmetry with respect to the underlying tetragonal crystal symmetry. In addition, in comparison with the chiral $p$-wave order, the nematic pairing could better explain the absence of split transitions under external perturbations that lift the degeneracy of the two $E_u$ components, as we shall explain later.

In a similar vein, odd-parity nematic superconductivity has been proposed in the doped topological insulator, Bi$_2$Se$_3$ \cite{Fu:10,Fu:14,Venderbos:16}, which has strong SOC and whose resultant rotational symmetry breaking has been reported in a few measurements \cite{Matano:16,Yonezawa:16,Pan:16,Du:16,Shen:17}.

{\it The Gingzburg-Landau theory}.---The generic two-component odd-parity superconducting pairing function in the $E_u$ representation reads,
\begin{equation}
\hat{\Delta}_{\bs k} =i( \phi_1 \vec{d}_{1,\bs k}\bs\cdot \vec{\sigma} + \phi_2 \vec{d}_{2,\bs k}\bs\cdot \vec{\sigma})\sigma_y \,.
\label{eq:gapSOC}
\end{equation}
where $\phi_{1,2}$ label the order parameters associated with the two components, $\vec{d}_{i,\bs k}$ are real vectors denoting the spin structure of the two components of Cooper pairing. The components of $\vec{d}_{i,\bs k}$ contain appropriate form factors [e.g. (\ref{eq:newD})] which form a two-dimensional $E_u$ representation of the underlying tetragonal crystalline space group $D_{4h}$ \cite{Volovik:85,Sigrist:91} . Throughout the work we assume only intraband Cooper pairing near the Fermi level, as is appropriate for a weak-coupling superconductor.

In the absence of SOC, the $\vec{d}$-vectors can be written in a separable form, $\vec{d}_{i,\bs k} = \vec{d}f_{i,\bs k}$, due to full spin rotational invariance. Here, the form factors $f_{i,\bs k}$ act as the basis functions of the corresponding symmetry group. In the presence of SOC,  Eq. (\ref{eq:gapSOC}) is more appropriately expressed in the pseudospin basis, whereby we would have duly accounted for the effects of SOC. In particular, since the spin and orbital degrees of freedom are now entangled, the $\vec{d}$-vector is locked with the momentum of the Bloch electron with pseudo-spin indices. In other words, the elements of the symmetry group operate simultaneously on the spin and the spatial coordinates.

We start our analyses with a phenomenological Gingzburg-Landau free energy. Up to the quartic order:
\begin{eqnarray}
f&=& r(T-T_c)\left(|\phi_1|^2 + |\phi_2|^2\right) + \beta \left(|\phi_1|^4+|\phi_2|^4 \right) \nonumber \\
&&+ \beta_{12}|\phi_1|^2|\phi_2|^2 + \beta^\prime \left(\phi_1^\ast\phi_2 + \phi_1\phi_2^\ast \right)^2 \,.
\label{eq:GLaction}
\end{eqnarray}
Within mean-field, coefficients of the quartic terms determine the stable superconducting state. In Ref. \cite{supp} we derive the expressions for evaluating these coefficients, from where it is apparent that $\beta$ and $\beta_{12}$ are positive definite. By contrast, the sign of $\beta^\prime$ depends on the structure of $\vec{d}_{i,\bs k}$ and can be roughly approximated by,
\begin{equation}
\beta^\prime \approx C \left\langle  (\vec{d}_{1,\bs k}\cdot\vec{d}_{2,\bs k})^2 - |\vec{d}_{1,\bs k} \times \vec{d}_{2,\bs k} |^2  \right \rangle_\text{FS} \,,
\label{eq:betaPtext}
\end{equation}
where $C$ is a positive constant and $\langle \cdots \rangle_\text{FS}$ denotes an integral over the Fermi surface. Similar expression was also obtained in Ref. \cite{Venderbos:16}. By inspection, if $\beta^\prime>0$, $\phi_1$ and $\phi_2$ preferentially develop a $\pi/2$ phase difference, i.e. $\bs{\phi}= (\phi_1,\phi_2)=\phi(1,\pm i)$, thereby breaking time-reversal invariance, as for the chiral $p$-wave order; whilst if $\beta^\prime<0$, the system favors a TRI order parameter, $\bs{\phi} = \phi(1, \pm 1)$ or $\phi(1,0)$. Note that the theory applies equally well to single- and multi-band models.

\begin{table}[t]
\caption{\label{tab:table1} Irreducible odd-parity representations of the $D_{4h}$ group and the corresponding basis functions.}
{\renewcommand{\arraystretch}{1.3}
\begin{tabular}{c|c}
\hline
~~~~irrep.~~~~ & ~~~~~~~~~basis function ($\vec{d}_{\bs k}$)~~~~~~~\\
\hline
$A_{1u}$                &     $k_x \hat{x} + k_y \hat{y}$; $k_z\hat{z}$  \\
\hline
$A_{2u}$                    &     $k_y \hat{x} - k_x \hat{y}$  \\ \hline
$B_{1u}$     &     $k_x\hat{x}-k_y\hat{y}$                \\ \hline
$B_{2u}$    &     $k_y\hat{x} + k_x \hat{y}$                                  \\ \hline
$E_{u}$       &     $(k_x,k_y)\hat{z}; ~~(k_z \hat{x}, k_z \hat{y})$ \\ \hline
\end{tabular}}
\label{tab:tab1}
\end{table}

{\it Interlayer-coupling-enforced 3D $\vec{d}$-vector}.--- Thus far, the assignment of the possible $\vec{d}$-vector in \SRO~has been largely dictated by the considerations of its layered structure. In particular, the quasi-2D character of the electronic structure naturally leads one to conjecture the absence of interlayer Cooper pairing that takes the form of, e.g. $\Delta_{\bs k} \sim k_z$ in the $p$-wave channel. However, no particular symmetry constraint prohibits such a pairing. Indeed, interlayer pairings of one form or another have been considered in a few microscopic models formulated in different contexts \cite{Hasegawa:00,Zhitomirsky:01,Annett:02}.

According to the classification in Table \ref{tab:tab1}, when a $k_z$-like pairing does develop, the superconducting state in the $E_u$ representation is more appropriately described by the following $\vec{d}$-vector,
\begin{equation}
\vec{d}_{1,\bs k} = k_x \hat z + \epsilon k_z \hat x ~~~~~\text{and}~~~~~ \vec{d}_{2,\bs k} = k_y \hat z+\epsilon k_z \hat y\,,
\label{eq:newD}
\end{equation}
where $\epsilon$ is a non-universal real constant determined by the relative strength of the effective interactions responsible for the respective $(k_x,k_y)\hat{z}$ and $(k_z\hat{x},k_z\hat{y})$ pairings, respectively. This alternative $\vec{d}$-vector texture was alluded in \cite{Kim:17}.

\begin{figure}
\includegraphics[width=8.cm]{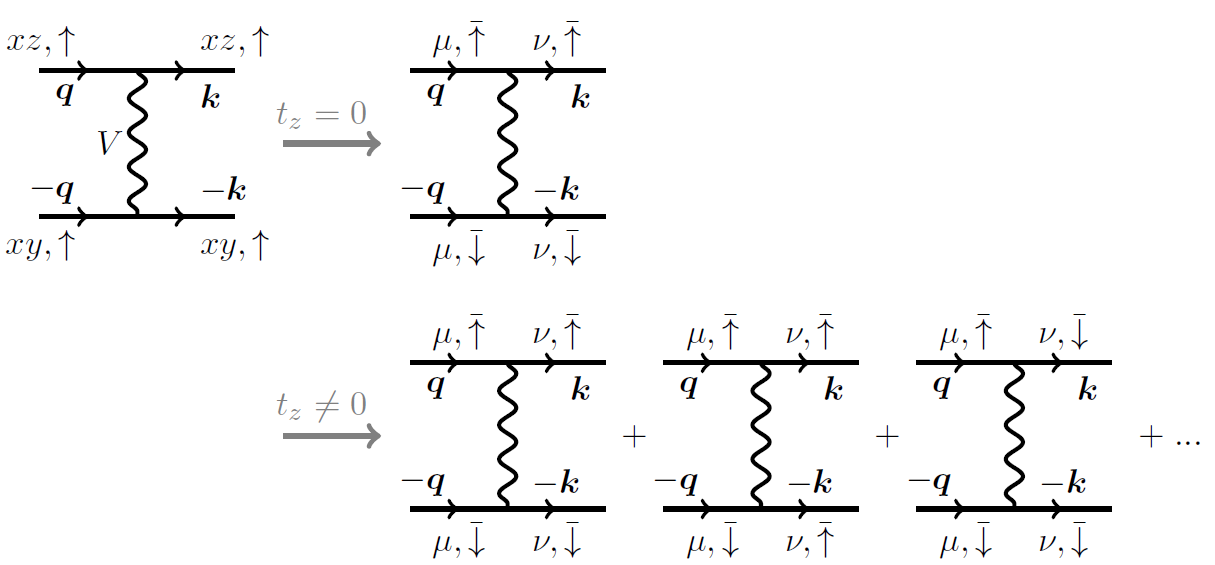}
\caption{Projection of a representative lowest order vertex into the pseudospin basis. The wavy lines denote the bare or projected Coulomb interactions. The diagram on the left originates from the bare Coulomb repulsion between the spin-up $xz$ and $xy$ electrons. The first and second line on the right hand side are for spin-orbit entangled three-orbital models without and with direct inter-orbital $xy$ - $xz/yz$ hopping $t_z$, respectively. The summation over the band indices $\mu/\nu$ are implicit. Note that only intraband pairing is considered. In addition, due to the peculiar form of the SOC, when $t_z=0$ the spin-up (down) $xy$-orbital is projected solely to the pseudospin-down (up) states in the band basis.}
\label{fig:feynman}
\end{figure}

Most previous quasi-two-dimensional spin-orbit coupled models of \SRO~do not support coexisting $(k_x,k_y)\hat{z}$ and $(k_z\hat{x},k_z\hat{y})$ pairings. This is due to the absence of direct inter-orbital hopping (or hybridization) between the $xy$ and the $xz/yz$-orbitals in those models. To understand this, we study the corresponding single-particle Hamiltonian,
\begin{equation}
H = \sum_{{\bs k},s} \psi^\dagger_{{\bs k},s} \hat{H}_{0s}({\bs k})\psi_{{\bs k},s} \,,
\label{eq:H0main}
\end{equation}
where the sub-spinor $\psi_{{\bs k},s} = (c_{xz,{\bs k},s},c_{yz,{\bs k},s},c_{xy,{\bs k},-s})^T$ with $c_{a,{\bs k},s}$ annihilating a spin-$s$ electron on the $a$-orbital ($a=xz,yz,xy$), $s=\ua~\text{and}~\da$ denote up and down spins, and,
\begin{equation}
\hat{H}_{0s}({\bs k}) = \begin{pmatrix}
\xi_{xz,\bs k}   &  \lambda_{\bs k} - is\eta  & i\eta  \\
\lambda_{\bs k} + is\eta  & \xi_{yz,\bs k}   &  -s\eta \\
-i\eta & -s\eta & \xi_{xy,\bs k}
\end{pmatrix} \,,
\label{eq:H0a}
\end{equation}
with $\xi_{a,\bs k}$ given in Ref. \cite{supp}. Here $\lambda_{\bs k}$ is the inter-orbital hybridization between the quasi-1D $xz$- and $yz$-orbitals; and $\eta$ is the strength of SOC. The eigenstates of (\ref{eq:H0main}) constitute the pseudospin electrons in the band representation. It is convenient to define the states associated with $\psi_{\bs k,\ua(\da)}$ pseudospin-up (down) particles. In this case, each pseudospin electron does not carry weight of more than one spin species of any orbital. Inversely, the decomposition of any individual spin species of any orbital belong with only one pseudospin species on the bands. Taking into account the on-site Coulomb interactions between the $t_{2g}$ orbitals, a low-energy effective action in the Cooper channel can be constructed perturbatively by projecting the Coulomb interactions and the associated spin/charge-fluctuation-mediated interactions onto the Fermi level \cite{Raghu:10,Scaffidi:14,Zhang:17b}.

\begin{figure}[t]
\includegraphics[width=8.cm]{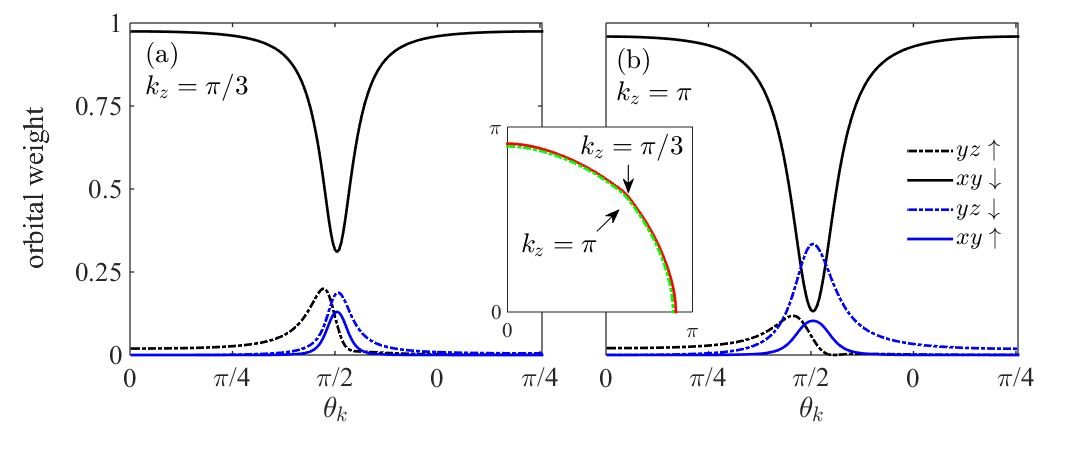}
\caption{Spin-resolved orbital weights across the $\gamma$-band Fermi surface at two different values of $k_z$ in the three-band tight-binding model with $t_z=0.05t$. The corresponding tight-binding model is given in Ref. \onlinecite{supp}. Presented data are for the pseudospin-up states as defined in the text. Only the weights of the $yz$ and $xy$-orbitals are shown for clarity. In the horizontal axis $\theta_k$ stands for the angle of the Fermi momentum w.r.t. the $x$-axis. Inset: Cross-section of the $\gamma$-band Fermi surface at the two $k_z$'s shown in (a) and (b).}
\label{fig:orbWeight}
\end{figure}

A close inspection of the projection reveals that the bare Coulomb interactions do not lead to scatterings from equal-pseudospin Cooper pairs to opposite-pseudospin pairs, nor vice versa. More specifically, as explained in Ref. \onlinecite{supp} and exemplified diagramatically in the first line of Fig \ref{fig:feynman}, interactions such as the following are absent at this order:
\begin{equation}
 \Gamma_{\bs k,\bs q} a^\dagger_{\nu,\bs k,\bar{\ua}}a^\dagger_{\nu,-\bs k,\bar{\ua}} a_{\mu,-\bs q,\bar{\ua}} a_{\mu,\bs q,\bar{\da}}
\label{eq:vertex}
\end{equation}
where $\Gamma_{\bs k,\bs q}$ denotes the projected effective interaction and $a_{\mu,\bar{\sigma}}^\dagger$ ($a_{\mu,\bar{\sigma}}$) creates (annihilates) a $\mu$-band pseudospin-$\sigma$ electron. Here the bar atop the spin symbol denotes pseudospin basis. It can be further verified that effective interactions like (\ref{eq:vertex}) remain absent even when higher order scattering processes are considered. As a consequence, the $x/y$ and $z$-components of the $\vec{d}$-vector of the concerned pseudospin-triplet channel are decoupled. Hence the $(k_x,k_y)\hat{z}$ and $(k_z\hat{x},k_z\hat{y})$ pairings in general need not coexist, or shall condense at different temperatures if they do at low-$T$.

However, in real material there always exists finite, albeit relatively weak, $xy$ and $xz/yz$ hybridization once interlayer coupling is considered. As is evident in Fig \ref{fig:orbWeight} and as explained in more detail in Ref. \onlinecite{supp}, even a relatively weak interlayper hopping $t_z$ between the $xy$ and $xz/yz$ orbitals could yield strong modifications to the details of the electronic structure. The resultant pseudospins, especially those with momenta around the Brillouin zone diagonal, possess sizable weights of both spin species of all orbitals. Remarkably, this is achieved without introducing noticeable corrugation to the three-dimensional Fermi surface (inset of Fig. \ref{fig:orbWeight}). We stress that the significant $k_z$-dependence of the spin-orbit entanglement was also observed in spin-resolved photo-emission studies \cite{Veenstra:14}. The corresponding action now permits scattering processes like (\ref{eq:vertex}), as illustrated in the second line of Fig. \ref{fig:feynman} (see also Ref. \onlinecite{supp}). In this case, the gap functions $(k_x,k_y)\hat{z}$ and $(k_z\hat{x},k_z\hat{y})$ are inherently coupled and should emerge simultaneously at a single $T_c$.

Note that although the realistic pairing functions are likely more anisotropic than shown in (\ref{eq:newD}), however, assuming the general relevance of the $(k_z\hat{x},k_z\hat{y})$ pairing, here we are only interested in the nontrivial consequences of the resultant 3D odd-parity pairing. 

\begin{figure}[t]
\includegraphics[width=5.cm]{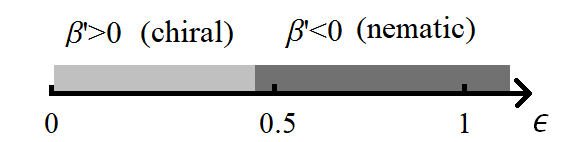}
\caption{The phase diagram as a function of $\epsilon$. The chiral and nematic phases are shaded in light and dark grey, respectively. We assume cylindrical Fermi surface with radius 1 and replace $k_z$ by $\sin k_z$ (taking the range of $k_z$ to be $[-\pi,\pi]$ in the integral). In this calculation $\beta^\prime$ changes sign at $\epsilon_c\approx 0.46$. }
\label{fig:phaseDiag}
\end{figure}

{\it Odd-parity nematic pairing}.---We proceed to discuss the stable superconducting states associated with (\ref{eq:newD}) in a single-band model for illustration. These $\vec{d}$-vectors lead to a simple expression for (\ref{eq:betaPtext}),
\begin{equation}
\beta^\prime \approx C\left\langle k_x^2k_y^2 - \epsilon^2k_z^2( k_x^2+k_y^2 +\epsilon^2 k_z^2)    \right\rangle_\text{FS} \,.
\label{eq:betaPnematic}
\end{equation}
Depending on the value of the anisotropic parameter $\epsilon$, $\beta^\prime$ can take either signs. In Fig. \ref{fig:phaseDiag}, we sketched the sign of $\beta^\prime$ as a function of $\epsilon$, assuming a cylindrical Fermi surface with radius $k_{F\parallel}=1$ and taking $k_z \ra \sin k_z$. Note that the critical value $\epsilon_c$ at which $\beta^\prime=0$ will in general be different in a more realistic model. As discussed, $\epsilon= \epsilon_c$ separates the TRSB and TRI phases. In the latter case, either a diagonal nematic state with $\bs \phi = \phi(1,\pm 1)$ or a horizontal one with $\bs \phi = \phi(1,0)/(0,1)$ could be more stable, depending on the details of the realistic band and gap structures \cite{supp}.

For small $|\epsilon|$, $\beta^\prime>0$ and the system favors a TRSB chiral-like pairing ($\epsilon=0$ returns the ordinary chiral $p$-wave) with a nodeless isotropic superconducting gap. This state is non-unitary, and it generates finite edge current. We do not further explore this possibility, but emphasize its intrinsic 3D nature if indeed realized in \SRO.

Below we focus on the TRI states, e.g. the horizontal nematic state $\bs \phi = \phi(1,0)$. Its gap function $|\Delta_{\bs k}|=\Delta_0 \sqrt{\epsilon^2k_z^2 + k_x^2 }$ reflects a breaking of the lattice $C_4$ symmetry down to $C_2$, with gap minima at $k_x=0$ and maxima at $k_x=\pm k_F$ at each $k_z$. Hence it may be termed a ``TRI nematic pairing". This state possesses nodal points at TRI $k_z$, i.e. $k_z=0$ and $k_z=\pi$ upon replacing $k_z$ by appropriate lattice harmonics in the gap function. Note that in other nematic states, the nodal-point directions are properly rotated. The representative gap structure at various $k_z$ is shown in Fig. \ref{fig:GapStruct} for the two different types of nematic states.

\begin{figure}[t]
\subfigure{\includegraphics[width=3.6cm]{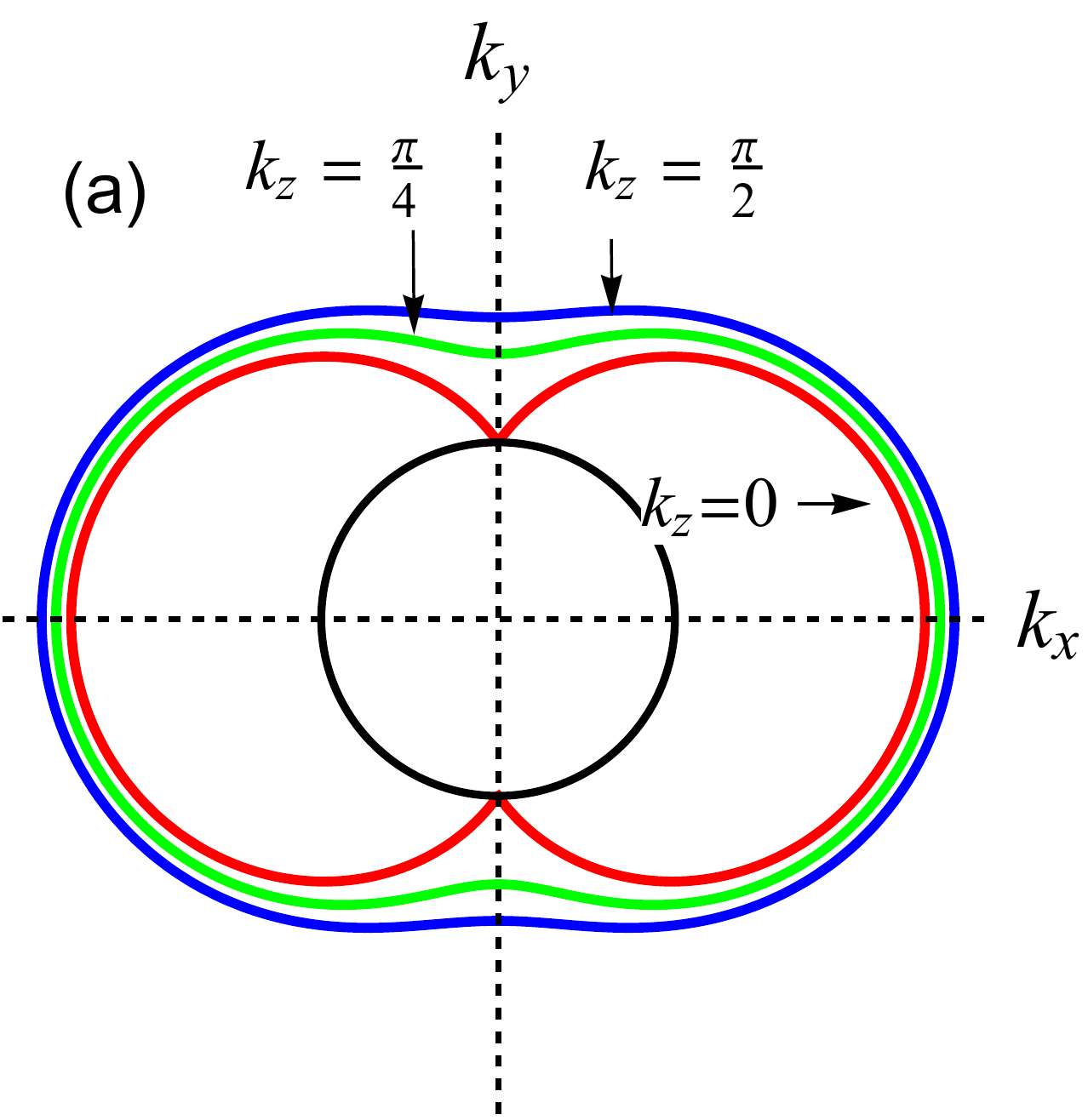} }
\subfigure{\includegraphics[width=3.5cm]{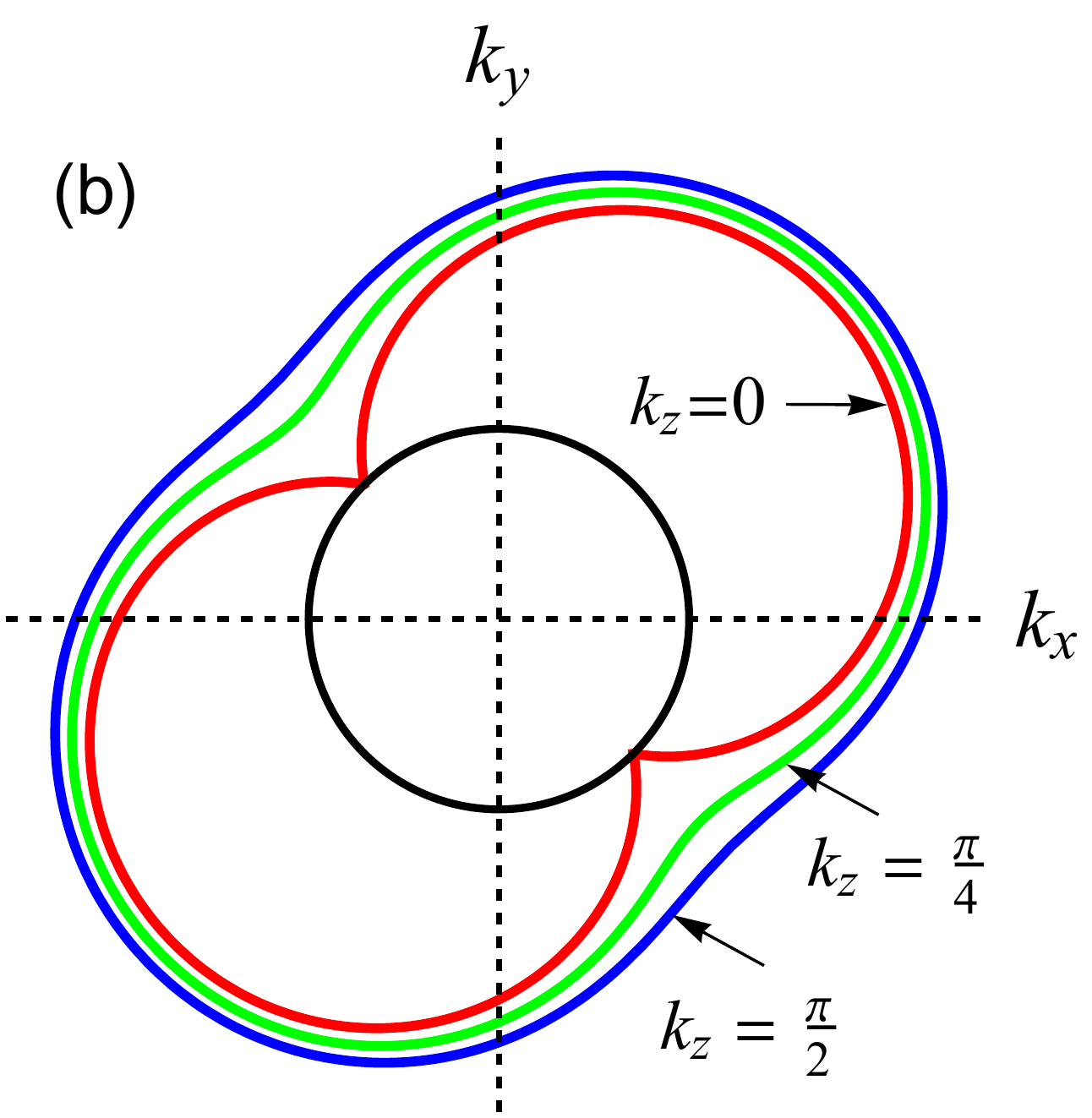} }
\caption{Representative gap structure on the approximately cylindrical Fermi surface (black) for the (a) horizonal nematic and (b) diagonal nematic states with $\epsilon =0.5$. As indicated, the three curves shown in each plot are for three values of $k_z$. In these calculations, we have replaced $k_z$ in the gap function by $\sin k_z$. }
\label{fig:GapStruct}
\end{figure}

One appealing feature of the nematic phase is the absence of spontaneous current at the edges due to TRI. In addition, at (100) or (010) surfaces, dispersionless edge modes with energy $ \propto \epsilon k_z$ emerge for each $k_z$ (Fig \ref{fig:EdgeModes}a) for the case of diagonal nematic pairing. Note that the horizontal nematic state with $\bs \phi \sim (1,0)$ supports surface states on the (100)-surface, but not on the (010)-surface. These surface states could be associated with the conductance peaks in the tunneling spectra \cite{Kashiwaya:11,Ying:12}. Taking the (100) surface in the diagonal nematic state as an example, there should be two horizontal Majorana arcs within $k_y \in [-k_{F\parallel}, k_{F\parallel}]/\sqrt{2}$ at $k_z=0$ and $\pi$ (Fig. \ref{fig:EdgeModes}b). By contrast, in the non-unitary chiral state, two singly-degenerate chiral edge modes appear (Fig.\ref{fig:EdgeModes}c) and finite edge current follows naturally. The zero modes in this case form two vertically connected and elongated Majorana loops, as depicted in Fig. \ref{fig:EdgeModes}c. The peculiar surface spectra of the nematic and chiral states can be distinguished in photoemission and quasiparticle interference studies. Furthermore, the nematic states form the bases of a $U(1) \times Z_2$ field theory. The $Z_2$ symmetry permits the formation of domains characterized by the two degenerate pairing functions, much like what has been proposed for chiral $p$-wave. This may be consistent with the signatures of domain formation observed in some experiments \cite{Kallin:09,Kidwingira:06,Saitoh:15,Wang:16,Anwar:17}.

{\it Discussions and summary}.---Guided by symmetry considerations, we argued that the odd-parity $E_u$ pairing in \SRO~acquires an inherent 3D form $(k_x\hat{z}+\epsilon k_z \hat{x}, k_y\hat{z}+\epsilon k_z\hat{y})$ in the presence of finite SOC and inter-orbital interlayer coupling. This leads to an appealing possibility of a novel TRI nematic odd-parity pairing, thereby providing an alternative perspective to understand the perplexing superconductivity in this material.

Besides resolving the notorious edge current problem, the nematic pairing may also explain some other outstanding puzzles. For example, compared with the chiral pairing, the nematic state could stand a better chance to explain the absence of split transitions under perturbations that break the degeneracy of the two $E_u$ components \cite{Yonezawa:14,Hicks:14,Steppke:16}. With applied uniaxial strain along any generic direction, the splitting is expected for the scenario with the chiral ground state where a sequence of two transitions spontaneously break distinct symmetries: the $U(1)$ symmetry at the upper transition, and the time-reversal symmetry at the lower one. For the case of diagonal nematic pairing, a genuine second transition occurs only when a uniaxial strain is applied {\it exactly} parallel to the (100)- or (010)-direction. In this scenario, the lower transition breaks a reflection symmetry about the vertical plane parallel to the (100)- or (010)-direction. However, for any amount of misalignment in applied strain, as could be the case in a real experiment, no sharp lower transition should occur \cite{footnoteIsing}, although a smeared crossover may appear to a degree that depends on the level of misalignment. Note also that, in the case of the horizontal nematic state, applying external strain along the (100)-direction cannot give rise to split transitions. 

\begin{figure}[t]
 \includegraphics[width=3.6cm,height=2.5cm]{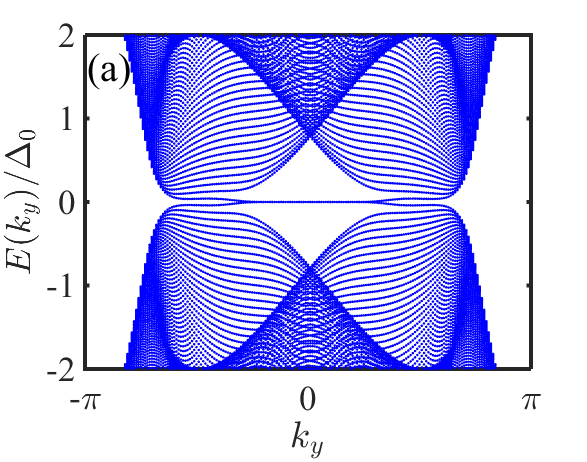}
 \includegraphics[width=4.9cm]{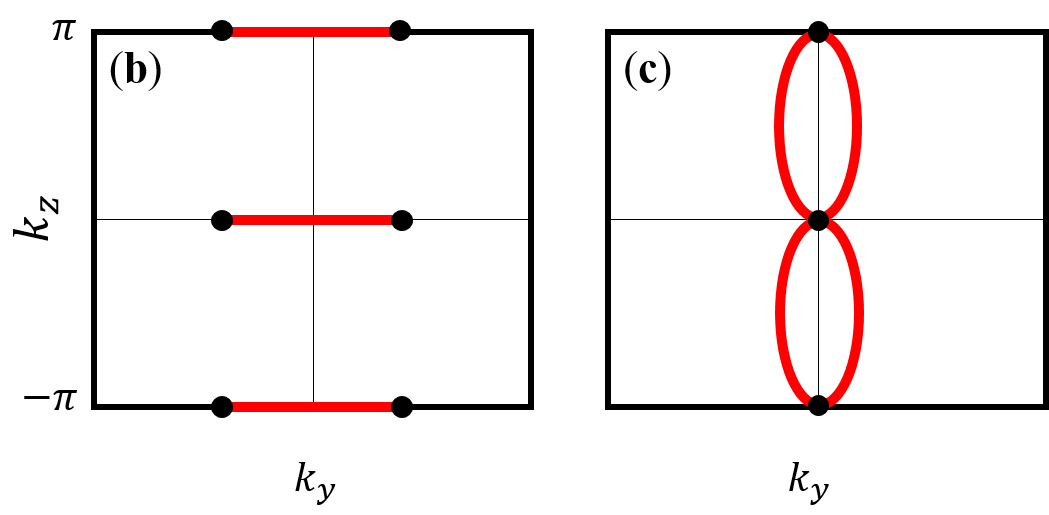}
\caption{(a) low-energy spectra at fixed $k_z=0$ for the diagonal nematic odd-parity pairing at $\epsilon=0.8$ in a geometry with two (100) surfaces. In the lattice BdG calculations, the $k_i$'s in the pairing functions are replaced by the simple lowest order harmonic $\sin k_i$. Note that the flat edge modes acquire some dispersion due to finite size effects. (b) sketched arcs spanned by the surface zero modes in the nematic phase. (c) sketched Majorana loops formed by the Majorana zero modes in the chiral phase.}
\label{fig:EdgeModes}
\end{figure}

Nevertheless, this nematic pairing needs to withstand the test of various other existing measurements, which remains to be carefully examined. Note that the absence of $C_2$ anisotropy in thermodynamic measurements under in-plane magnetic fields \cite{Mao:00,Deguchi:04} could be consistent with nematic pairing because the externally applied field may drive a rotation of the nematic orientation. By contrast, in Bi$_2$Se$_3$ superconductors the nematic orientation may be pinned by a weak structural distortion as reported in Ref. \onlinecite{Kuntsevich:18}. The nematicity in \SRO~could be revealed in measurements like the angle-dependent in-plane Josephson tunneling \cite{Strand:10} and the visualization of single vortex structure in scanning tunneling microscopy, as has been demonstrated for Cu$_x$Bi$_2$Se$_3$ \cite{Tao:18,footnoteVortex}. Regarding the contradiction between the point-nodal gap structure and the experimental indications of line-nodal pairing \cite{Nishizaki:00,Hassinger:16}, we note that the realistic gap function could be more anisotropic. In particular, it is in principle possible to have, e.g. $\vec{d}_{1\bs k} = g_{\bs k}(k_x\hat{z}+\epsilon k_z \hat{x})$ and similarly for $\vec{d}_{2\bs k}$, where the form factor $g_{\bs k}$ carries horizontal or vertical line nodes. Such line nodes are not imposed by symmetry but could very likely arise in reality, given the highly anisotropic electronic structure in \SRO.

{\it Acknowledgements}. We are grateful to Fan Yang and Li-Da Zhang for stimulating discussions, and to Daniel Agterberg and Shingo Yonezawa for helpful communications. This work is supported in part by the MOST of China under Grant Nos. 2016YFA0301001 and 2018YFA0305604 and by the NSFC under Grant No. 11474175 (W.H. and H.Y.). W.H. also acknowledges the support by the C. N. Yang Junior Fellowship of the Institute for Advanced Study at Tsinghua University.


\widetext
\begin{center}
\section{ Supplemental Materials for ``Possible three-dimensional nematic odd-parity superconductivity in \SRO"}
\end{center}
\setcounter{equation}{0}
\setcounter{figure}{0}
\setcounter{table}{0}
\makeatletter
\renewcommand{\theequation}{S\arabic{equation}}
\renewcommand{\thefigure}{S\arabic{figure}}
\renewcommand{\bibnumfmt}[1]{[S#1]}

\author{Wen Huang}
\affiliation{Institute for Advanced Study, Tsinghua University, Beijing, 100084 China}
\author{Hong Yao}
\affiliation{Institute for Advanced Study, Tsinghua University, Beijing, 100084 China}
\affiliation{State Key Laboratory of Low Dimensional Quantum Physics, Tsinghua University, Beijing 100084, China}

\date{\today}

\maketitle

\section*{I. Effective two-dimensional three-band models}
We first present the effective multi-orbital model, constructed from the three Ru $t_{2g}$ orbitals, which is commonly employed in previous studies of \SRO, and is a purely two-dimensional model without interlayer coupling. By symmetry direct inter-orbital $xy$ - $xz/yz$ hopping (hybridization) is absent, but the sizable spin-orbit coupling can be accounted for from the outset. Repeating the main text for completeness, the single particle Hamiltonian reads,
\begin{equation}
H = \sum_{{\bs k},s} \psi^\dagger_{{\bs k},s} \hat{H}_{0s}({\bs k})\psi_{{\bs k},s} \,,
\label{eq:H0}
\end{equation}
where the sub-spinor $\psi_{{\bs k},s} = (c_{xz,{\bs k},s},c_{yz,{\bs k},s},c_{xy,{\bs k},-s})^T$ with $c_{a,{\bs k},s}$ annihilating a spin-$s$ electron on the $a$-orbital ($a=xz,yz,xy$), $s=\ua~\text{and}~\da$ denote up and down spins, and,
\begin{equation}
\hat{H}_{0s}({\bs k}) = \begin{pmatrix}
\xi_{xz,\bs k}   &  \lambda_{\bs k} - is\eta  & i\eta  \\
\lambda_{\bs k} + is\eta  & \xi_{yz,\bs k}   &  -s\eta \\
-i\eta & -s\eta & \xi_{xy,\bs k}
\end{pmatrix} \,,
\label{eq:H0aA}
\end{equation}
with $\xi_{xz,\bs k} = -2t \cos k_x -2 \tilde{t} \cos k_y -\mu$, $\xi_{yz,\bs k} = -2\tilde{t} \cos k_x -2 t \cos k_y-\mu$, $\lambda_{\bs k} =-$2$t^{\prime\prime} \sin k_x \sin k_y $, $\xi_{xy,\bs k} = -2t^\prime(\cos k_x + \cos k_y) -4t^{\prime\prime\prime} \cos k_x \cos k_y -\mu_1$. Here $\lambda_{\bs k}$ is the inter-orbital hybridization between the two quasi-1D $xz$- and $yz$-orbitals, and $\eta$ is the strength of spin-orbit coupling (SOC). In the main text we use the set of parameters $(\tilde{t},t^\prime,t^{\prime\prime},t^{\prime\prime\prime},\eta,\mu_,\mu_1)=(0.1,0.8,0.1,0.3,0.1,1,1.1)t$, which well reproduces the band structure and the Fermi surface geometry of \SRO. Note that because SOC mixes different spins on the $xy$- and the other two orbitals, the spins are not good quantum numbers. However, thanks to the inversion symmetry the Kramers degeneracy on each band is preserved, it is therefore convenient to adopt a pseudospin notation and denote the degenerate electrons pseudospin-up and down. In this case, the two pseudospin species are fully characterized by the respective sub-spinors $\psi_{\bs k,s}$, e.g. the pseudospin-up electron is linearly composed only of $xz \ua$, $xz \ua$ and $xy \da$ electrons but carries not weight of $xz \da$, $xz \da$ and $xy \ua$. Similar statements can be made in reverse. For example, the $xz \ua$, $yz \ua$ and $xy \da$ electrons can only be decomposed into pseudo-spin up electrons in the band basis.

To understand the projection of the pairing vertex onto the Fermi level, we again look at the example given in the main text, i.e. with the bare inter-orbital Coulomb interaction between $xz \ua$ and $xy \ua$ electrons:
\begin{eqnarray}
Vn_{xz,\ua}n_{xy,\ua} &\ra& V c^\dagger_{xz,\bs k,\ua}c^\dagger_{xy,-\bs k,\ua} c_{xy,-\bs q,\ua} c_{xz,\bs q,\ua} \nonumber \\
&=& \sum_{\mu,\nu} \Gamma_{xz \ua,xy \ua;xy \ua,xz \ua}^{\nu \bar{\ua},\nu \bar{\da};\mu \bar{\da},\mu \bar{\ua}} (\bs k, \bs q) a^\dagger_{\nu,\bs k,\bar{\ua}}a^\dagger_{\nu,-\bs k,\bar{\da}} a_{\mu,-\bs q,\bar{\da}} a_{\mu,\bs q,\bar{\ua}}\, ,
\label{eq:Vdecomp}
\end{eqnarray}
with the effective pairing vertex given by,
\begin{equation}
\Gamma_{xz \ua,xy \ua;xy \ua,xz \ua}^{\nu \bar{\ua},\nu \bar{\da};\mu \bar{\da},\mu \bar{\ua}} (\bs k, \bs q) = V \left[\xi_{xz\ua}^{\nu\bar{\ua}} (\bs k) \xi_{xy\ua}^{\nu\bar{\da}} (-\bs k)\right]^\ast \xi_{xy\ua}^{\mu\bar{\da}} (-\bs p) \xi_{xz\ua}^{\mu\bar{\ua}} (\bs p) \,.
\label{eq:Vdecomp1}
\end{equation}
Here the coefficients $\xi_{a\sigma}^{\mu\bar{\sigma}^\prime}(\bs k)$ are the corresponding elements of the unitary transformation relating the orbital and band representations. Note in the above expressions we have used the fact that $\xi_{xz\ua}^{\mu(\nu)\bar{\da}}=\xi_{xy\ua}^{\mu(\nu)\bar{\ua}} \equiv 0$ when the $xy$ - $xz/yz$ hybridization is absent. It then follows that effective interactions such as (7) in the main text cannot appear in the low-energy theory. This can be shown to be true for all other bare Coulomb interactions and higher-order scatterings. As a side remark, only intraband pairing is considered under the weak-coupling assumption in the present study.


\section*{II. Role of $xy$ - $xz/yz$ hopping (hybridization)}
A more complete description of the electronic structure necessarily involves nonvanishing $xy$ - $xz/yz$ hopping (between same spin species). To make explicit its influence to the electronic structure, we consider the lowest order contribution which involves interlayer coupling. We expand the full single-particle Hamiltonian (\ref{eq:H0aA}) in the matrix form as follows,
\begin{equation}
H = \sum_{{\bs k}} \Psi^\dagger_{{\bs k}} \hat{H}_{0}({\bs k})\psi_{{\bs k}} \,,
\label{eq:H0full}
\end{equation}
where the full spinor $\Psi_{{\bs k}} =(\psi_{\bs k,\ua}^T,\psi_{\bs k,\da}^T)= (c_{xz{\bs k}\ua},c_{yz{\bs k}\ua},c_{xy{\bs k}\da},c_{xz{\bs k}\da},c_{yz{\bs k}\da},c_{xy{\bs k}\ua})^T$ and,
\begin{equation}
\hat{H}_{0}({\bs k}) = \begin{pmatrix}
\xi_{xz,\bs k}   &  \lambda_{\bs k} - i\eta  & i\eta   &   0    &   0    &    T_{1z}(\bs k)\\
\lambda_{\bs k} + i\eta  & \xi_{yz,\bs k}   &  -\eta &  0    &   0    &    T_{2z}(\bs k) \\
-i\eta & -\eta & \xi_{xy,\bs k}   &   T_{1z}(\bs k)   &    T_{2z}(\bs k)   &  0 \\
0    &   0    &    T^\ast_{1z}(\bs k) & \xi_{xz,\bs k}   &  \lambda_{\bs k} + i\eta  & i\eta   \\
0    &   0    &    T^\ast_{2z}(\bs k) &   \lambda_{\bs k} - i\eta  & \xi_{yz,\bs k}   &  \eta \\
T^\ast_{1z}(\bs k)   &    T^\ast_{2z}(\bs k)   &  0 & -i\eta & \eta & \xi_{xy,\bs k}
\end{pmatrix} \,.
\label{eq:H0afull}
\end{equation}
Here $T_{iz}(\bs k)$ represents the $xy$ - $xz/yz$ hopping terms. To leading order, they may be approximated by $T_{1z}(\bs k) = 8 t_z \cos\frac{k_x}{2} \sin\frac{k_y}{2} \sin \frac{k_z}{2}$ and $T_{2z}(\bs k) = 8 t_z \sin\frac{k_x}{2} \cos\frac{k_y}{2} \sin \frac{k_z}{2}$ where $t_z$ is the corresponding interlayer hopping amplitude between the 1D and 2D orbitals. Notably, since the interlayer hopping preserves the inversion symmetry, the Kramers degeneracy remains. Hence the notion of pseudospins remains valid. When these interlayer hoppings are absent, the Hamiltonian returns to (\ref{eq:H0aA}) and is block-diagonalized.

This leads to a finite corrugation of the 3D Fermi surface along the $k_z$-axis. As can be inferred from the inset of Fig 2, the corrugation is minuscule for a relatively weak $t_z$, as is consistent with the experiments. In spite of this, the influence on the pseudospin structure is noticeable. This is evident from the sizable mixture of spin and orbital species belonging to the two spinor subspace, which is not accessible in models with vanishing $xy$ - $xz/yz$ hopping. Due to the level-crossing of the unhybridized orbital dispersions around the BZ diagonals, the mixing is strongest in these regions. The $k_z$-dependence of the pseudospin structure is also evident in Fig 2. Note that since $T_{iz}(\bs k)$ vanishes at $k_z=0$, the mixture of the two sub-spinors $\psi_{\bs k,s}$ is suppressed at this $k_z$.

The projection of the interactions follows similar procedure as in (\ref{eq:Vdecomp}). However, due to the mixing of all orbital and spin species at $k_z \neq 0$, the example given in the previous section now permits effective vertices $\Gamma$ with generically finite $\xi_{xz\ua}^{\mu(\nu)\bar{\da}}$ and $\xi_{xy\ua}^{\mu(\nu)\bar{\ua}}$. As a consequence, effective interactions such as (7) in the main text are generally allowed in the effective action.

As a side remark, besides making the $E_u$-pairing inherently three-dimensional, the same mechanism also makes the $A_{u}$ state three-dimensional, namely, $\vec{d}_{\bs k} = k_x \hat{x} + k_y\hat{y} + \epsilon k_z \hat{z}$ according to the classification in Table I, which is similar to the B-phase of $^3$He. In addition to being fully-gapped, this state is singly-degenerate, hence no domain formation is expected. 

\section*{III. Ginzburg-Landau $\beta$-coefficients}
For a generic two-component odd-parity pairing function $\hat{\Delta}_{\bs k} =i( \phi_1 \vec{d}_{1,\bs k}\bs\cdot \vec{\sigma} + \phi_2 \vec{d}_{2,\bs k}\bs\cdot \vec{\sigma})\sigma_y $, the free energy functional is obtained via a standard perturbative expansion in powers of the order parameter fields $\phi_1$ and $\phi_2$. The Gorkov Greens function in Nambu spinor space reads,
\begin{equation}
\hat{G}^{-1}(i w_n, \bs k) = \begin{pmatrix}
(iw_n -\xi_{\bs k}) \sigma_0 & \hat{\Delta}_{\bs k} \\
\hat{\Delta}_{\bs k}^\dagger &(iw_n +\xi_{\bs k})\sigma_0
\end{pmatrix} \,,
\end{equation}
where $\sigma_0$ is the rank-2 identity matrix, $w_n=(2n+1)\pi T$ is the Matsubara frequency with $T$ near $T_c$, around which the expansion is valid. Defining $g_+(iw_n,\bs k) = \frac{1}{iw_n-\xi_{\bs k}}$, $g_-(iw_n,\bs k) = \frac{1}{iw_n+\xi_{-\bs k}}$,  $\hat{G}_0^{-1} = \begin{pmatrix} g_+^{-1} \sigma_0 & 0 \\ 0 & g_-^{-1} \sigma_0  \end{pmatrix}$ and $\hat{\Delta} = \begin{pmatrix} 0 & \hat{\Delta}_{\bs k} \\ \hat{\Delta}_{\bs k}^\dagger & 0  \end{pmatrix}$, the part of the expansion essential to our discussion reads,
\begin{eqnarray}
-T\cdot \text{tr~ln}\hat{G}^{-1} &=& -T \cdot\text{tr~ln}(\hat{G}_0^{-1}+\hat{\Delta}) = -T \cdot\text{tr~ln}\hat{G}_0^{-1}(1+\hat{G}_0\hat{\Delta}) \nonumber \\
&=&  \text{const.}-T\cdot \text{tr~ln}(1+\hat{G}_0\hat{\Delta})  \nonumber \\
&=& \text{const.} + \frac{T}{2l}\sum_{l=1}^{\infty} \text{tr} [\hat{G}_0\hat{\Delta}]^{2l} \,\,.
\label{eq:Expansion1}
\end{eqnarray}
Note only even-order contributions survive in the last line. The $l=1$ term yields,
\begin{eqnarray}
&& \frac{T}{2} \text{tr} \left[ \hat{G}_0 \hat{\Delta} \hat{G_0} \hat{\Delta}   \right]  = \frac{T}{2L^d}\sum_{w_n,\bs k} g_+g_- \text{tr}\left[ \hat{\Delta}_{\bs k} \hat{\Delta}_{\bs k}^\dagger \right] \nonumber \\
&&= \frac{T}{2L^d}\sum_{w_n,\bs k} g_+g_- \text{tr}\left[ (\phi_1 \vec{d}_{1\bs k} \cdot \vec{\sigma}+ \phi_2 \vec{d}_{2\bs k}\cdot \vec{\sigma}) (\phi_1^\ast \vec{d}_{1\bs k}\cdot \vec{\sigma} + \phi_2^\ast \vec{d}_{2\bs k}\cdot \vec{\sigma}) \right] \nonumber \\
&&= \frac{T}{2L^d}\sum_{w_n,\bs k} g_+g_- \text{tr}\left[ |\vec{d}_{1\bs k}|^2 |\phi_1|^2 \sigma_0 + |\vec{d}_{2\bs k}|^2 |\phi_2|^2  \sigma_0  +(\vec{d}_{1\bs k}\cdot\vec{\sigma}) (\vec{d}_{2\bs k}\cdot\vec{\sigma}) \phi_1^\ast\phi_2 + (\vec{d}_{2\bs k}\cdot\vec{\sigma}) (\vec{d}_{1\bs k}\cdot\vec{\sigma}) \phi_2^\ast \phi_1  \right] \,.
\label{eq:Expansion2}
\end{eqnarray}
Noting that $(\vec{d}_{1\bs k} \cdot \vec{\sigma})(\vec{d}_{2\bs k} \cdot \vec{\sigma})= (\vec{d}_{1\bs k}\cdot \vec{d}_{2\bs k})\sigma_0 + i \vec{\sigma}\cdot(\vec{d}_{1\bs k}\times \vec{d}_{2\bs k})$, and that the last two terms in the last line can be shown to vanish upon $\bs k$-summation. Therefore $\phi_1$ and $\phi_2$ do not couple at this quadratic order. Proceeding to the quartic order, $l=2$, using (\ref{eq:Expansion2}),
\begin{eqnarray}
&& \frac{T}{4} \text{tr} \left[ \hat{G}_0 \hat{\Delta} \hat{G_0} \hat{\Delta}   \right]^2  = \frac{T}{4L^d}\sum_{w_n,\bs k}(g_+g_-)^2 \text{tr}\left[ \hat{\Delta}_{\bs k} \hat{\Delta}_{\bs k}^\dagger \right]^2 \nonumber \\
&&= \frac{T}{4L^d}\sum_{w_n,\bs k}(g_+g_-)^2 \text{tr}\left\{ |\vec{d}_{1\bs k}|^2 |\phi_1|^2  \sigma_0+ |\vec{d}_{2\bs k}|^2 |\phi_2|^2  \sigma_0 + (\vec{d}_{1\bs k}\cdot \vec{d}_{2\bs k})\sigma_0 (\phi_1^\ast\phi_2 + \phi_2^\ast \phi_1) + i \vec{\sigma}\cdot(\vec{d}_{1\bs k}\times \vec{d}_{2\bs k}) (\phi_1^\ast\phi_2 - \phi_2^\ast \phi_1)  \right\}^2 \,. \nonumber \\
&&
\label{eq:Expansion3}
\end{eqnarray}
Expanding the curly bracket we obtain the quartic terms. By comparing with (2) in the main text, it is readily seen that,
\begin{eqnarray}
\beta &=& \frac{T}{4L^d}\sum_{w_n,\bs k}(g_+g_-)^2 |\vec{d}_{1\bs k}|^2  ~\text{tr}( \sigma_0)  = \frac{T}{2L^d}\sum_{w_n,\bs k} \frac{1}{(w_n^2+\xi_{\bs k}^2)^2} |\vec{d}_{1,\bs k}|^4  \nonumber \\
&\propto& \left\langle |\vec{d}_{1,\bs k}|^4 \right\rangle_\text{FS}   \,,
\label{eq:beta}
\end{eqnarray}
\begin{equation}
\beta_{12} = \frac{T}{2 L^d}\sum_{w_n,\bs k} \frac{1}{(w_n^2+\xi_{\bs k}^2)^2} \left[ |\vec{d}_{1,\bs k}|^2 |\vec{d}_{2,\bs k}|^2  + 2|\vec{d}_{1,\bs k} \times \vec{d}_{2,\bs k} |^2 \right] \times 2
\label{eq:beta12}
\end{equation}
\begin{equation}
\beta^\prime= \frac{T}{2L^d}\sum_{w_n,\bs k} \frac{1}{(w_n^2+\xi_{\bs k}^2)^2} \left[(\vec{d}_{1,\bs k}\cdot\vec{d}_{2,\bs k})^2 - |\vec{d}_{1,\bs k} \times \vec{d}_{2,\bs k} |^2 \right] \,.
\label{eq:betaP}
\end{equation}
The $\langle ... \rangle_\text{FS}$ in the second line of (\ref{eq:beta}) denotes a Fermi surface integral. Similar, although not exactly equivalent (due to a different form of free energy functional we adopt), expressions were obtained in Ref. [74]. Note that after completing the Matsubara frequency summation, the $\bs k$-summation can be approximated as an average across the Fermi surface, as examplified in (\ref{eq:beta}). In addition, terms like $|\phi_1|^2(\phi_1^\ast \phi_2 + \phi_2^\ast\phi_1)$ vanish, as their coefficients do not survive the $\bs k$-summation.

\section*{IV. Horizontal Vs diagonal nematic pairing}
We consider the scenario where the nematic pairing is favored, i.e. $\beta^\prime<0$. Rewriting the free energy functional (2) in the main text as follows,
\begin{equation}
f=r(T-T_c) (|\phi_1|^2+|\phi_2|^2) + \beta (|\phi_1|^2 + |\phi_2|^2)^2 + (\beta_{12}-2\beta+4\beta^\prime)|\phi_1|^2|\phi_2|^2+\beta^\prime(\phi_1^\ast\phi_2 - \phi_2^\ast\phi_1)^2 \,,
\label{eq:GLsupp}
\end{equation}
it is clear that, up to the quartic order, if $\beta_{12}-2\beta+4\beta^\prime=0$, the states with $(\phi_1,\phi_2)=\phi_0 (\cos \theta, \sin\theta)$ are degenerate for all nematic $\theta$, exhibiting a continuous degeneracy. This continuous degeneracy is present if the system respects the in-plane rotational symmetry. However, real materials possess only discrete lattice rotational symmetry. As a result, $\beta_{12}-2\beta+4\beta^\prime\neq 0$ and the continuous degeneracy is generically lifted. Through straightforward saddle point approximation, it can be shown that the diagonal nematic pairing with $\phi_0(1,\pm 1)/\sqrt{2}$ is preferred when $\beta_{12}-2\beta+4\beta^\prime < 0$, otherwise the horizontal nematic phase with $\phi_0(1, 0)/(0,1)$ is more stable. As can be inferred from the previous section, the choice of nematic angle is determined by the details of the realistic band and gap structures.


\begin{thebibliography}{99}
\bibitem{footnoteIsing} This is similar to the scenario in the classical Ising model, where the transition is smeared by any magnetic field except those applied precisely perpendicular to the Ising spins.
\bibitem{footnoteVortex} The scanning SQUID can also image the vortex structure \cite{Hicks:10}. However, unlike STM, the current scanning SQUID measurements may not have the precision to resolve the nematic vortex shape.
\end{thebibliography}
\end{document}